\pdfoutput=1

\documentclass[apj]{emulateapj}
\slugcomment{}

\shorttitle{SFHs of low-mass star forming galaxies}
\shortauthors{Rodr\'{\i}guez-Mu\~noz et al.}

\begin{document}

\title{RECENT STELLAR MASS ASSEMBLY OF \\LOW-MASS STAR FORMING GALAXIES AT REDSHIFTS 0.3~$<$~\MakeLowercase{\emph{z}}~$<$~0.9\footnotemark[*]}\footnotetext[*]{B\MakeLowercase{ased on observations carried out with the }E\MakeLowercase{uropean} S\MakeLowercase{outhern} O\MakeLowercase{bservatory} (ESO) V\MakeLowercase{ery} L\MakeLowercase{arge} T\MakeLowercase{elescope} (VLT) \MakeLowercase{at the} L\MakeLowercase{a} S\MakeLowercase{illa} P\MakeLowercase{aranal} O\MakeLowercase{bservatory under programs 088.}A-\MakeLowercase{0321 and 090.}A-\MakeLowercase{0858.}}

\author{Luc\'{\i}a Rodr\'{\i}guez-Mu\~noz\altaffilmark{1}, Jes\'{u}s Gallego\altaffilmark{1}, Camilla Pacifici\altaffilmark{2}, Laurence Tresse\altaffilmark{3}, \\St\'{e}phane Charlot\altaffilmark{4}, Armando Gil de Paz\altaffilmark{1}, Guillermo Barro\altaffilmark{5}, and V\'{\i}ctor Villar\altaffilmark{1}}

\affil{\altaffilmark{1} Dpto. de Astrof\'isica y CC. de la Atm\'osfera. Facultad de CC. F\'isicas, Universidad Complutense de Madrid, \\Av. Complutense s/n. E-28040, Madrid, Spain}
\email{lucia.rodriguez.munoz@fis.ucm.es}

\affil{\altaffilmark{2} Yonsei University Observatory, Yonsei University, Seoul 120-749, Korea}

\affil{\altaffilmark{3} Aix Marseille Universit\'e, CNRS, LAM (Laboratoire d'Astrophysique de Marseille) UMR 7326, 13388, Marseille, France}

\affil{\altaffilmark{4} UPMC-CNRS, UMR7095, Institut d'Astrophysique de Paris, F-75014 Paris, France}

\affil{\altaffilmark{5} UCO/Lick Observatory and Department of Astronomy and Astrophysics, University of California, Santa Cruz, CA 95064, USA}

\begin{abstract}
The epoch when low-mass star forming galaxies (LMSFGs) form the bulk of their stellar mass is uncertain.
While some models predict an early formation, others favor a delayed scenario until later ages of the universe. 
We present constraints on the star formation histories (SFHs) of a sample of LMSFGs obtained through the analysis of their 
spectral energy distributions (SEDs) using a novel approach that: (1) consistently
combines photometric (broadband) and spectroscopic (equivalent widths
of emission lines) data, and (2) uses physically motivated SFHs with
non-uniform variations of the star formation rate (SFR) as a function of time. 
The sample includes 31 spectroscopically confirmed LMSFGs (7.3~$\le$~log$M_{*}/M_{\sun}$~$\le$~8.0), at 0.3~$<$~$z_{\mathrm{spec}}$~$<$~0.9, in the Extended-Chandra Deep Field-South field. 
Among them, 24 were selected with photometric stellar mass log$M_{*}/M_{\sun}$~$<$~$8.0$, $0.3$~$<$~$z_{\mathrm{phot}}$~$<$~$1.0$, and $m_{\mathrm{NB816,AB}}$~$<$~$26$~mag; the remaining 7 were selected as blue compact dwarfs within the same photometric redshift and magnitude ranges. We also study a secondary sample of 43 more massive spectroscopically confirmed galaxies (8.0~$<$~log$M_{*}/M_{\sun}$~$\le$~9.1), selected with the same criteria. The SFRs and stellar masses derived for both samples place our targets on the standard main sequence of star forming galaxies. The median SFH of LMSFGs at intermediate redshifts appears to form 90\% of the median stellar mass inferred for the sample in the $\sim$~0.5--1.8~Gyr immediately preceding the observation. These results suggest a recent stellar mass assembly for LMSFGs, consistent with the cosmological downsizing trends. We find similar median SFH timescales for the more massive secondary sample.
\end{abstract}

\keywords{galaxies: dwarf -- galaxies: evolution -- galaxies: formation -- galaxies: star formation}

\section{Introduction}
Galaxy formation and evolution is a key area of current research in extragalactic astrophysics \citep[e.g.,][]{2012opsa.book...39B}. 
In the last decade, an important step forward has been taken due to the development of models and simulations, and the impact of deep multiwavelength photometric and spectroscopic surveys. 
While the history of low-mass dark matter halos is theoretically well 
understood, models and simulations still struggle to reproduce observations due 
to the distinct evolution of baryonic and dark
matter \citep{2012AN....333..515S}. 

Low-mass galaxies appear as one of the most poorly known systems. How and when these objects assemble their mass is still a source of debate. Early models \cite[e.g.,][]{1986ApJ...303...39D} predicted that low-mass galaxies should experience the bulk of their star formation (SF) before the end of reionization, $z$\,$\sim$\,6 \citep[]{2013ASSL..396...45Z}. Later, \cite{1997ApJ...487...61K} proposed a scenario for these systems where the SF is delayed until $z$\,$\sim$\,1. More recently, \cite{2012dgkg.book...39M} proposed a mass-dependent theoretical scheme, in which only extremely low-mass galaxies (log$M_{*}/M_{\sun}$~$\lesssim$~7) form before the end of reionization. 

The most detailed studies of star formation histories (SFHs) of low-mass galaxies have been carried out in the local universe \citep[for example, works from the Local Cosmology from Isolated Dwarfs project, e.g.,][]{2011ApJ...730...14H}. Although every system appears to present an old stellar population, observational studies disagree about when they undergo their dominant star forming activity. For instance, \citet{2011ApJ...739....5W} found an early epoch of formation prior to $z$\,$\sim$\,1 of 60\% of the stellar masses of a morphologically heterogeneous sample of 60 Local Group low-luminosity/mass galaxies, analyzing the color-magnitude diagrams of their resolved stellar populations. In contrast, \cite{2012ApJ...745..149L} inferred a late ($z$\,$\leq$\,1) assembly of 70\% of the stellar mass for low-mass star forming galaxies (LMSFGs) with masses $8.0$~$<$~log$M_{*}/M_{\sun}$~$<$~$8.5$ using both a main sequence integration (MSI) approach and the spectral energy distribution (SED) fitting of a sample of star forming galaxies from the Sloan Digital Sky Survey \citep[SDSS;][]{2002AJ....124.1810S}. 

The importance of studying isolated and higher redshift low-mass galaxies has been demonstrated \citep[e.g.,][]{2012dgkg.book....3S}. In this way, recent works have been carried out to constrain the properties of low-mass field galaxies along cosmic timescales \citep[e.g.,][]{2013ApJ...776L..27H,2014A&A...568L...8A,2014ApJ...788L...4A,2014arXiv1403.3441A}. These studies have used different selection criteria to build their samples of targets (e.g. color, luminosity, extreme nebular emission).  In our study, we do not select our sample using features that probably vary on short timescales, but we use a more stable parameter such as the stellar mass. Thus, we have in hand a sample that is less biased toward certain evolutionary stages. 

Our objective is to constrain the SFHs of LMSFGs in order to reduce the uncertainties in their formation epoch, which is when they form the bulk of their stellar mass. We 
use an approach proposed by \cite{2012MNRAS.421.2002P} to study the SEDs of our sample of spectroscopically confirmed 
intermediate redshift field LMSFGs. This approach presents the novelty of considering physically motivated SF and chemical enrichment histories derived from semi-analytic models, instead of crude approximations based on simple functions. Moreover, it allows for the simultaneous study of the stellar and nebular emission by using both photometric and spectroscopic data. 

Throughout this paper we adopt a standard $\Lambda$-CDM cosmology with $H_{0}$\,$=$\,0.7, $\Omega_{m}$\,$=$0.3 and $\Omega_{\Lambda}$\,$=$\,0.7, AB~magnitudes and Chabrier initial mass function \citep[][]{2003PASP..115..763C}.

\section{Sample \& Data}

We build our sample of LMSFG candidates from a deep SUBARU NB816 image \citep{2006ApJ...638..596A} of the Extended-Chandra Deep Field-South \citep[E-CDF-S;][]{2005ApJS..161...21L} using the Rainbow database\footnote[1]{http://rainbowx.fis.ucm.es} \citep{2008ApJ...675..234P,2011ApJS..193...13B,2011ApJS..193...30B}. The catalog on this image is the deepest available in the visible wavelength range in Rainbow at the time the selection is made. The exhaustive deep multiband photometry from UV to far-IR over the E-CDF-S field enables the estimation of photometric redshifts and stellar masses through the analysis of their SEDs. Rainbow is also a template-fitting code based on $\chi^{2}$ minimization between observed photometry and a set of $\sim$\,1500 semi-empirical template SEDs (see \citet{2008ApJ...675..234P}, Appendix A). E-CDF-S has available morphology catalogs developed by \cite{2012ApJS..200....9G}. 

To identify candidates for low-mass galaxies, we use two samples selected by different criteria. On the one hand, a \emph{mass-selected sample} of $\sim$\,700 galaxies is built using the Rainbow photometric stellar masses and photometric redshifts: log$M_{*}/M_{\odot}$~$<$~8, and 0.3~$<$~$z_{\mathrm{phot}}$~$<$~1.0. We choose the upper mass value in order to identify a hypothetical bimodal behavior of the formation redshifts, as found by \cite{2012dgkg.book...39M}. Such value corresponds also to the range of halo galaxy masses expected to dominate the reionization of cosmic hydrogen \citep{2006ApJ...646..696W}. On the other hand, a \emph{blue compact dwarf (BCD) sample} \citep{1981ApJ...247..823T} is selected in the same photometric redshift range, using as selection criteria the following definition of BCD galaxies based on the work by \cite{2003ApJS..147...29G}: $M_{\mathrm{B},0,\mathrm{AB}}$~$>-18$~mag, 
$(B-V)_0$~$<$~0.6, and $S_{\mathrm{Reff,B,}0}$~$<$~$23$~mag\,arcsec$^{-2}$ . Among dwarfs, BCDs present observational advantages 
such as strong emission lines, resulting from the SF burst they undergo, and high surface brightness, which makes them excellent tracers of LMSFGs at intermediate redshifts. We found $\sim$\,$900$ galaxies matching these selection criteria. This sample includes only galaxies with stellar masses log$M_{*}/M_{\odot}$~$>$~8. 
We consider $\mathrm{m}_{\mathrm{NB816,AB}} < 26$ mag as an additional photometric criterion to target candidates susceptible of being spectroscopically confirmed using the VIsible Multi-Object Spectrograph \citep[VIMOS;][]{2003SPIE.4841.1670L}, mounted on the 
$8\,\mathrm{m}$ ESO-VLT/Unit Telescope 3 at the La Silla Paranal Observatory.

Subsequently, we perform VIMOS deep spectroscopy of a subsample of 327~candidates (253~low-mass selected and 74~selected as BCDs) during a Visitor Mode run and a Service 
Mode run in 2011 November and 2012 December, respectively (ESO programs 088.A-0321 and 090.A-0858.). The total number of targets observed is determined by the total exposure time granted and the slit masks design performed with the VIMOS Mask Preparation Software \citep[VMMPS;][]{2005PASP..117..996B}. This software places slits on the targets in a way that the final number of slits in each mask is maximized. We assign a higher priority to the mass-selected sample than to the BCDs sample.
The two runs use the medium resolution (MR) grism combined with GG475 filter. The width of the slits is 1$^{\prime\prime}$.2 and 1$^{\prime\prime}$.6 for the first and the second program, respectively. These configurations yield a spectral resolution $R\,\sim$\,600 and an effective spectral range 5000--10000~\AA. Observations are organized for three VIMOS pointings with total exposure times of 3.3, 2.7, and 3.8~hr, respectively.  

Data are reduced with the VIMOS Interactive Pipeline and Graphical Interface software \citep[VIPGI;][]{2005PASP..117.1284S} in combination with {\sc{reduceme}} software \citep{cardiel99}. VIPGI undertakes standard processing 
of bias subtraction and wavelength calibration of MR spectra, 
identification of objects in each slit profile, extraction of the one-dimensional
spectra, and flux calibration, all of which are optimized for VIMOS data. 

The wavelength calibration is carried out based on the arc lamp exposures obtained 
immediately following the science images, with a typical accuracy $\leq$\,0.3~pixel 
($\sim$\,1\AA). The flux calibration is carried out using flux standard stars observed following the default procedure for VIMOS observations, and average sensitivity curves provided by ESO. The calibration procedures are probed to be compatible within 20\% comparing galaxies common to both programs. 

We measure reliable spectroscopic redshifts, $z_{\mathrm{spec}}$, for those galaxies that present a minimum of two recognizable spectral features. The typical uncertainties for $z_{\mathrm{spec}}$ are around~0.1\%. The measurements of the emission line fluxes and equivalent widths (EW) are performed using {\sc{reduceme}} software \citep{cardiel99}. The galaxies mainly present strong emission lines such as [OII]3727\AA, H$\beta$, [OIII]4960,5007\AA, and H$\alpha$, but also other emission lines such as H3835\AA, [NeIII]3869\AA, H3889\AA, [NeIII]3968\AA, H$\delta$, H$\gamma$, [OIII]4363\AA, HeI\,4472\AA, HeI\,5876\AA, [NII]6583\AA, HeI\,6678\AA, [SII]6716\AA, [SII]6731\AA.

In this paper, we focus on the sample of 94 emission line galaxies (62 originally selected by stellar mass and 32 selected as BCDs) for which we find a reliable redshift. The values measured range between 0.1~$<$~$z_{\mathrm{spec}}$~$<$~1.3. The observed targets for which no redshift is measured ($\sim$70\% of the sample) do not present recognizable spectral features. Most of them show only an extremely faint continuum. The fact that we can only measure reliable spectroscopic redshifts for emission line galaxies introduces a bias in the sample selection toward star forming galaxies. Future publications (L. Rodr\'{\i}guez-Mu\~noz et al. in preparation) will present the data and the complete study of the properties of the  sample. 
 
The primary selection performed with Rainbow uses photometry based on the average ordinary galaxy population. In our case, a more careful photometric extraction is needed for the specific galaxies studied in this analysis. Hence, we perform reliable aperture photometry in different optical and near to medium IR bands across the E-CDF-S field using the Rainbow software package. In particular we use \emph{Hubble Space Telescope} (\emph{HST})/Advanced Camera for Surveys (ACS) $b$, $v$, $i$, $z$ bands, and deep VIMOS U and R images from the Great Observatories Origins Deep Survey \citep[GOODS;][]{2004ApJ...600L..93G, 2009ApJS..183..244N}, the Multiwavelength Survey by Yale-Chile (MUSYC) 18 medium-band imaging \citep{2010ApJS..189..270C}, and HST/Wide Field Camera 3 (WFC3) F105W, F145W, and F160W from the Cosmic Assembly Near-infrared Deep Extragalactic Legacy Survey \citep[CANDELS;][]{2011ApJS..197...35G,2011ApJS..197...36K}. For \emph{Spitzer}/InfraRed Array Camera (IRAC) 3.6, 4.5, 5.8, and 8.0~$\mu$m bands we use the mean of the value re-measured on the images from GOODS, and those available in the CANDELS-TFIT \citep{2013ApJS..207...24G} and MUSYC \citep{2010ApJS..189..270C} catalogs. In the end, we have a total of 39 photometric bands available. Typical uncertainties in the absolute photometric calibration are lower than 0.05 mag in all bands, as determined in \cite{2010ApJS..189..270C} for the MUSYC data (including small zeropoint offsets based on the comparison with templates used in the photometric redshift determination), \cite{2005PASP..117..978R} for the IRAC data, and \cite{2005PASP..117.1049S} and \cite{2009wfc..rept...30K} for the ACS/WFC3 data.

\section{Modeling \& Fitting Approach}

\begin{figure*}[th!]
   \centering
   \includegraphics[angle=0,width=\textwidth]{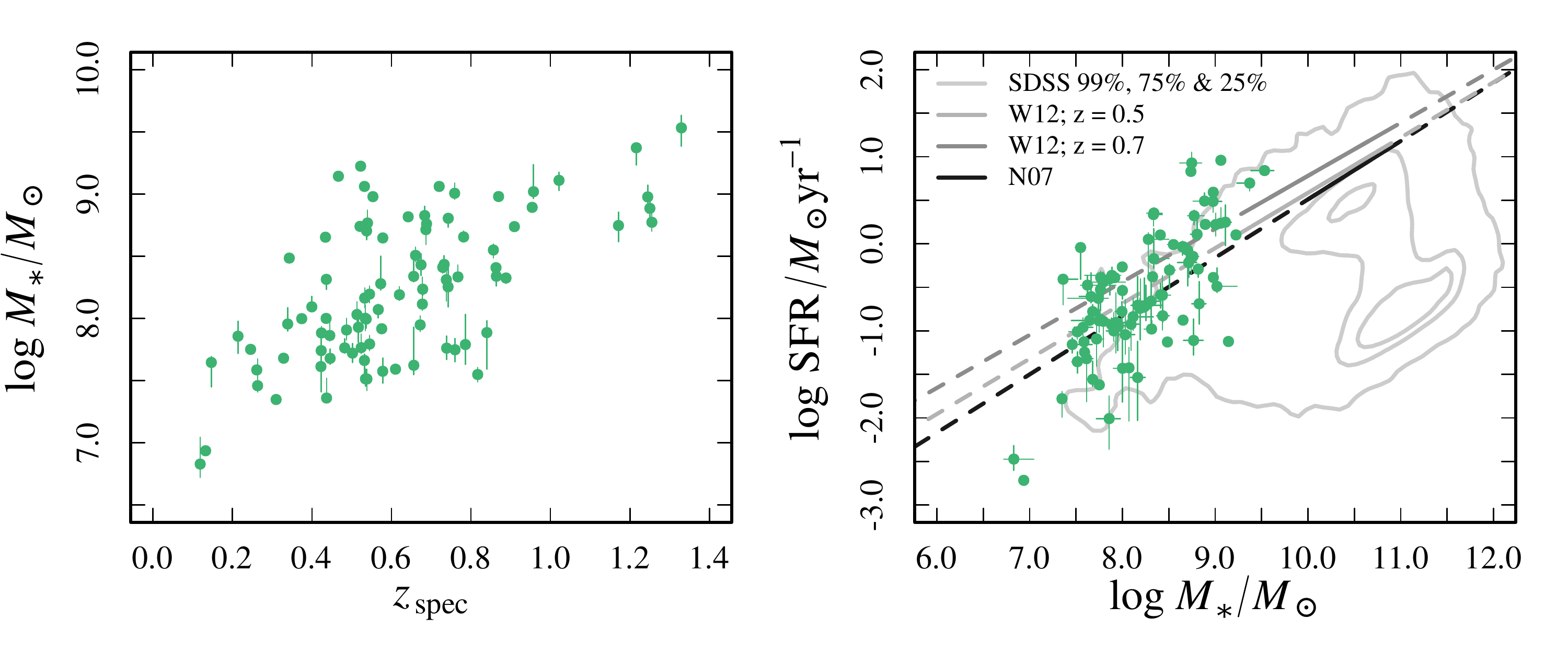}
   \caption{Left: stellar mass vs. redshift for our sample of 91 star forming galaxies. Right: stellar mass--SFR relation. The gray lines represent the MS by \citet[][N07]{2007ApJ...660L..43N}, and the MS derived by \citet[][W12]{2012ApJ...754L..29W} for redshifts 0.5 and 0.7. We mark with solid lines the stellar mass ranges within the mass-completeness limits of the studies. We use dashed lines to extend the MSs toward higher and lower stellar masses. In contours we show the distribution of the SDSS Data Release 7 \citep[][]{2009ApJS..182..543A}}. 
              \label{MS}%
    \end{figure*}

\begin{deluxetable*}{cccccc}
\label{table1}
\tablecaption{Main Properties of the Final Samples.}
\tablewidth{0pt}
\tablehead{
\colhead{Sample} & \colhead{\#} & \colhead{$z_{\mathrm{spec}}$} & \colhead{log$M_{*}/M_{\odot}$} & \colhead{logSFR} & \colhead{logsSFR}\\
\colhead{(1)} & \colhead{(2)} & \colhead{(3)} & \colhead{(4)} & \colhead{(5)} & \colhead{(6)}}
\startdata
LMSFGs & 31 & 0.517 (0.374, 0.672) & 7.7 (7.5, 7.9) & $-0.9$ ($-1.3$, $-0.4$) & 0.4 (0.0, 0.7)\\ 
Secondary & 43 & 0.656 (0.521, 0.743) & 8.4 (8.2, 8.8) & $-0.4$ ($-1.0$, 0.1) & 0.0 ($-0.5$, 0.3)\\
\enddata
\tablecomments{For each sample, we report: (1) name; (2) number of galaxies; (3) spectroscopic redshift; (4) stellar mass ($M_{\odot}$); (5) star formation rate ($M_{\odot}$yr$^{-1}$); (6) specific star formation rate (Gyr$^{-1}$); Columns (3--6) show median values, and 16\,th \& and 84\, percentiles (within parenthesis) of the distributions.}
\end{deluxetable*}

We use the spectral analysis tool developed by \cite{2012MNRAS.421.2002P} to obtain stellar masses and derive constraints on the SFHs of our sample. This tool combines a large library of physically motivated SF and chemical enrichment histories from cosmological simulations with state-of-the-art models of stellar population synthesis, nebular emission, and attenuation by dust, to provide us with best estimates and confidence ranges of stellar~mass~($M_{*}/M_{\odot}$), star formation rate (SFR, $M_{\odot}$yr$^{-1}$), specific star formation rate (sSFR~$=$~$\mathrm{SFR}/M_{*}$, Gyr$^{-1}$), along with a best-estimate SFH for each galaxy in the sample, using a Bayesian approach.

In practice, we build a large library of physically motivated SF and chemical enrichment histories applying semi-analytic recipes \citep{2007MNRAS.375....2D} to the output of the Millennium cosmological simulation \citep{2005Natur.435..629S}. To characterize the SFHs we define $t_{0}$ as the lookback time at the onset of SF, and $t_{10}$ and $t_{50}$ as the lookback times when a galaxy forms the 10 and 50\% of the total stellar mass. The library is tuned to avoid possible biases in the statistics and in the subsequent fitting process by (1) uniformly covering the ranges of physical parameters expected for our sample of LMSFGs, and (2) randomly drawing the evolutionary stages linearly in lookback time allowing for a uniform distribution of timescales ($t_{0}$) at each redshift.

We obtain a library of approximately 1.5 million spectral models by combining the SF and chemical enrichment histories library with state-of-the-art models of stellar population synthesis \citep[latest version of][]{2003MNRAS.344.1000B}, nebular emission based on the photoionization code CLOUDY \citep{1996hbic.book.....F} as in \cite{2001MNRAS.323..887C}, and the attenuation by dust model \`a la \cite{2000ApJ...539..718C}, as in \cite{2013ApJ...762L..15P}, \citep[see also][]{2013MNRAS.432.2061C}. In our case, we draw the total optical depth of the dust ($\hat{\tau}_{V}$) in the range $0$~$<$~$\hat{\tau}_{V}$~$<~1$ (instead of $0$~$<$~$\hat{\tau}_{V}$~$<~3$) as this is more suitable to fit our sample of blue LMSFGs.

Finally, we fit the observational data (photometry\footnote[2]{In the fitting procedure, we impose a minimum of 0.05 mag for the photometric uncertainties to enlarge the number of models contributing to each fit.} and EWs of the aforementioned emission lines) of the sample of 94~galaxies (Section~2) to the same observable quantities inferred for those models with redshifts within $z_{\mathrm{spec}}$~$\pm$~0.02 (in practice, each target is compared to about 65,000 models). We build a probability density function for the value of each parameter based on the likelihood of the fits. Median values of such probability distributions are recorded as the best-estimates of the parameters. The typical uncertainties (median half the 16--84\,th percentile range) are $\sim$\,$0.1$~dex for the stellar mass, $\sim$\,$0.1$~dex for the SFR, and $\sim$\,$0.2$~dex for the sSFR. Furthermore, a best-estimate SFH is derived as the average of the first 10 best-fit model SFHs weighted by their likelihood. Since it is very challenging to accurately constrain the full SFH of individual galaxies, in this paper we focus on average SFHs. 

We obtain fits for 91 galaxies. We exclude one galaxy with less than 12 photometric points and other 2 for which the fitting process returns near-zero probability for more than 95\% of the models in the library. The 91 galaxies present stellar masses between 6.8~$<$~log$M_{*}/M_{\sun}$~$<$~9.5. The differences between the Rainbow stellar masses used for the primary selection and the values obtained with this new approach present a significant dispersion with a median absolute deviation and 16\,th and 84\,th percentiles of 0.0, $-0.9$ and 0.4 respectively, but no systematic effects (Rodr\'{\i}guez-Mu\~noz et al. in preparation). For this reason no bias is introduced in the properties of the sample.

\section{Star Formation Histories}

In Figure~\ref{MS} we show the final sample of 91 star forming galaxies: The left panel shows the distribution of stellar mass with redshift. Our observational strategy causes a lack of low-mass objects at high redshifts. 
The right panel shows the stellar mass--SFR relation for our final sample. We note that the most massive galaxies tend to present higher values of SFR than the least massive ones, which is in agreement with studies about the main sequence (MS) of star forming galaxies \citep[e.g.,][]{2007ApJ...660L..43N,2012ApJ...754L..29W}. We limit our study to the redshift range 0.3~$<z<$~0.9, where the observed galaxies span roughly a uniform range in stellar mass ($7.3$~$\le$~log$M_{*}/M_{\sun}$~$\le$~$9.1$).

\begin{figure*}[t!]
   \centering
   \includegraphics[angle=0,width=\textwidth]{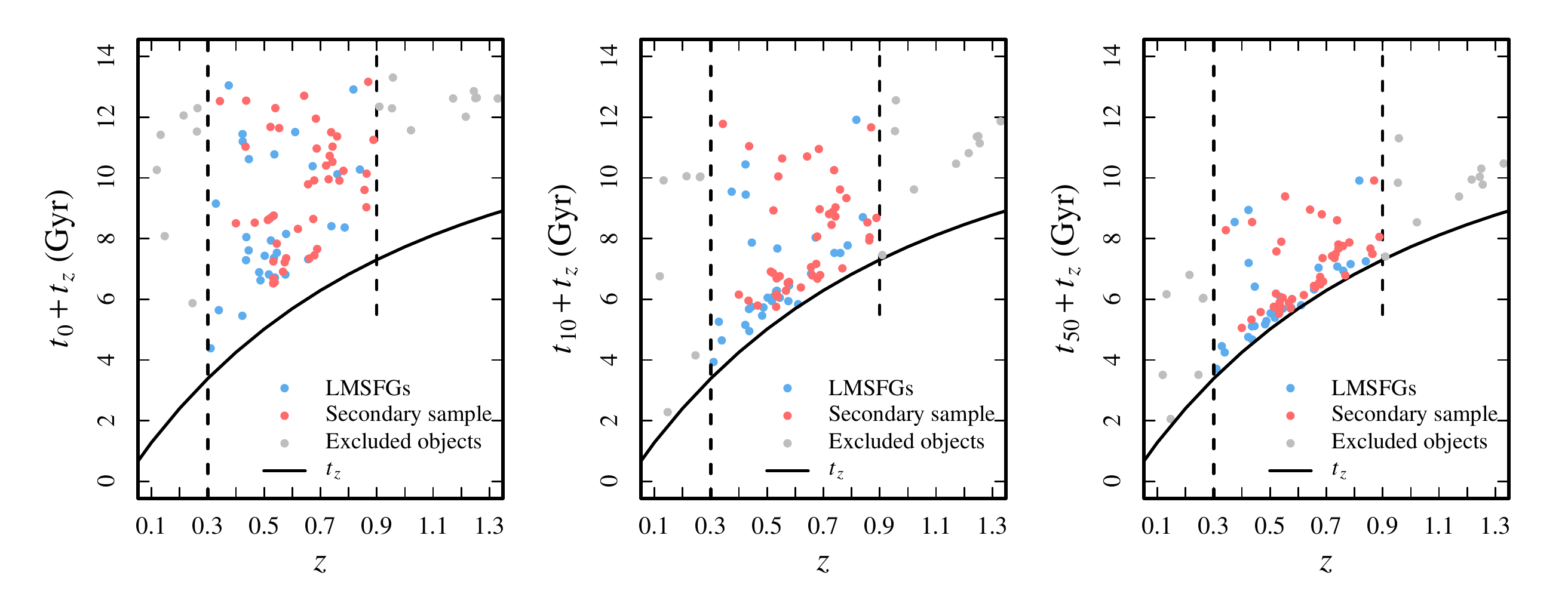}
   \caption{From left to right: $t_{0}$, $t_{10}$, and $t_{50}$ of each individual galaxy accounting for the lookback time at the redshift of observation ($t_{z}$, black solid line). The LMSFGs and the secondary sample are represented with blue and red points respectively. The gray points show the galaxies outside the final redshift range considered, marked by vertical black dashed lines.} 
              \label{ts}%
    \end{figure*}

We find 74 galaxies out of 91 (Section~3) within these redshift limits. We divide this final sample into two groups,
   \begin{enumerate}
      \item \emph{LMSFGs sample} (blue points in Figure~\ref{ts}): 31~LMSFGs with stellar masses between $7.3$~$\le$~log$M_{*}/M_{\sun}$~$\le$~$8.0$ (24 selected by the photometric stellar mass criterion and 7 selected as BCDs). 
      \item \emph{Secondary sample} (red points in Figure~\ref{ts}): 43~more massive galaxies, with stellar masses between $8.0$~$<$~log$M_{*}/M_{\sun}$~$\le$~$9.1$ (21 originally selected by the photometric stellar mass criterion and 22 selected as BCDs). 
   \end{enumerate}   
Table~1 gives the median values and 16\,th--84\,th percentiles of the stellar mass, SFR, and the sSFR for both final samples. 

\begin{deluxetable*}{cccccccccc}
\label{table2}
\tablecaption{SFHs Timescales of the Final Samples.}
\tablewidth{0pt}
\tablehead{
\colhead{Sample} & \colhead{}& \colhead{SFH-P50} &\colhead{} & \colhead{}& \colhead{SFH-P16} &\colhead{} & \colhead{}& \colhead{SFH-P84} & \colhead{}\\ \cline{2-4} \cline{5-7}\cline{8-10}
\colhead{} & \colhead{$t_{0}$} & \colhead{$t_{10}$} & \colhead{$t_{50}$} & \colhead{$t_{0}$} & \colhead{$t_{10}$} & \colhead{$t_{50}$} & \colhead{$t_{0}$} & \colhead{$t_{10}$} & \colhead{$t_{50}$}\\
\colhead{(1)} & \colhead{(2)} & \colhead{(3)} & \colhead{(4)} & \colhead{(5)} & \colhead{(6)} & \colhead{(7)} & \colhead{(8)} & \colhead{(9)} & \colhead{(10)} }
\startdata
LMSFGs & 2.5$\pm0.4$ & 0.8$\pm0.1$ & 0.3$\pm0.0$ & 1.3$\pm0.2$ & 0.5$\pm0.2$ & 0.3$\pm0.1$ & 5.0$\pm1.0$ &  1.8$\pm1.0$ & 0.5$\pm0.1$ \\
Secondary & 3.6$\pm0.3$ & 1.4$\pm0.3$ & 0.5$\pm0.1$ & 1.7$\pm0.3$ & 0.9$\pm0.1$ & 0.4$\pm0.1$ & 5.9$\pm0.7$ & 3.4$\pm0.7$ & 0.8$\pm0.2$ \\
\enddata
\tablecomments{For each sample, we report: (1) name; (2--4) $t_{0}$, $t_{10}$, and $t_{50}$ (Gyr) for SFH-P50; (5--7) $t_{0}$, $t_{10}$, and $t_{50}$ (Gyr) for SFH-P16; (8--10) $t_{0}$, $t_{10}$, and $t_{50}$ (Gyr) for SFH-P84; We present the average and standard deviations of the $t_{0}$, $t_{10}$, and $t_{50}$ of the SFH-P50, SFH-P16, and SFH-P84 obtained for 10$^{3}$ different bootstrap samples of SFHs drawn from each mass bin.}
\end{deluxetable*}

Figure~\ref{ts} shows the distribution in redshift of $t_{0}$, $t_{10}$, and $t_{50}$ mass assembly milestones. In each plot, the black solid line shows the lookback times corresponding to each redshift ($t_{z}$). Uncertainties for $t_{0}$, $t_{10}$ and $t_{50}$ are difficult to derive because the resolution of the SFH models decreases with lookback time (0.10--0.25~Gyr), and uncertainties of the stellar masses should also be taken into account. The age of the universe constrains the maximum lengths of the possible SFHs at each redshift. 

\begin{figure*}[t]
   \centering
   \includegraphics[angle=0,width=\textwidth]{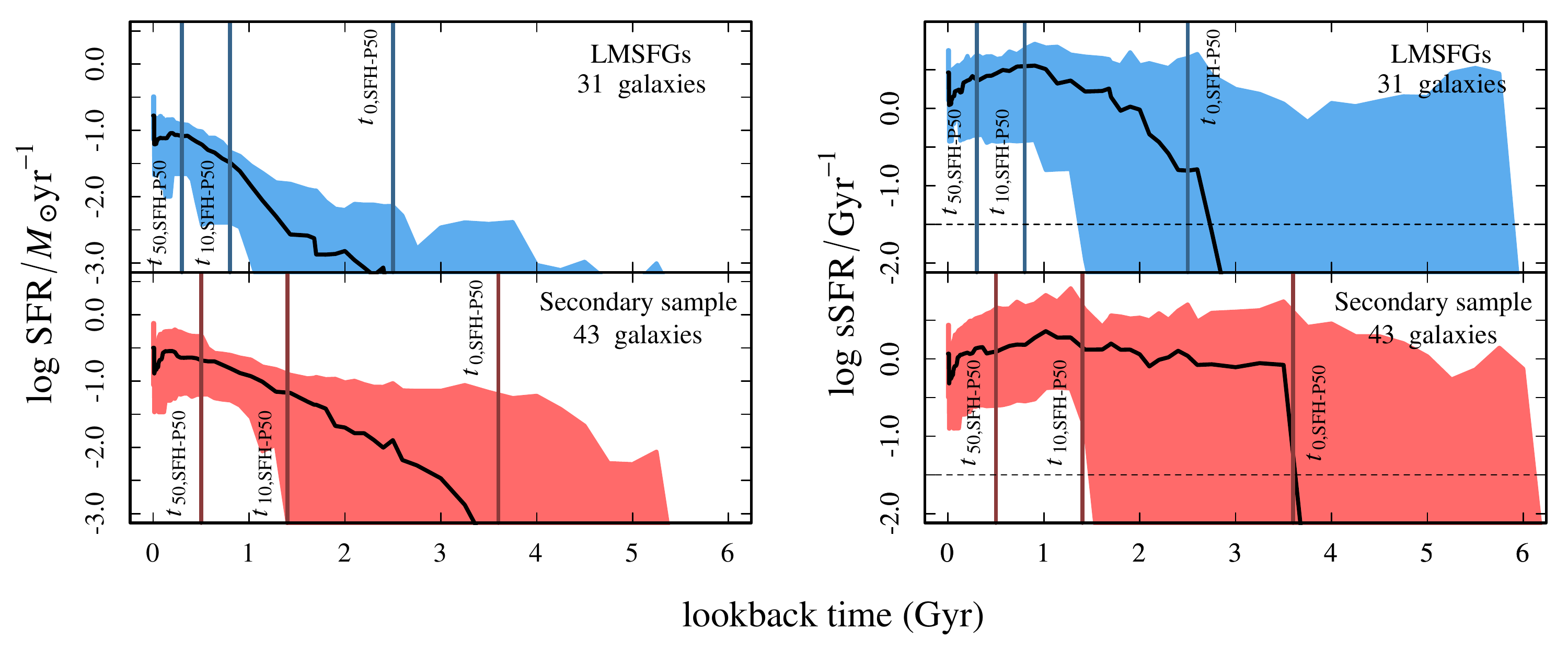}
   \caption{Left: median SFHs of the final samples (SFH-P50, black solid line). Individual SFHs are normalized to the median mass of the corresponding bin before being combined. The colored areas are delimited by the corresponding 16\,th and 84\,th percentiles of the SFR per lookback time step (SFH-P16 and SFH-P84, respectively). Right: median sSFRHs. The horizontal dashed lines mark the threshold between star forming and quiescent galaxies as in \cite{2009MNRAS.394.1131K}. In each panel, the vertical lines indicate the lookback time of the first SF burst of the SFH-P50 corresponding to each stellar mass bin ($t_{0,\mathrm{SFH-P50}}$), and when SFH-P50 forms 10\% and 50\% of the final stellar mass ($t_{10,\mathrm{SFH-P50}}$ and $t_{50,\mathrm{SFH-P50}}$). The number of galaxies included in each stellar mass bin is also indicated.\\ \\ \\}
              \label{Average}%
    \end{figure*}

In order to identify common behaviors, we stack the individual SFHs that best reproduce the observations in each final sample. In practice, (1) we normalize the SFHs to the median stellar mass of the corresponding sample; (2) we set each SFH to a common reference system where $t_{z}$~$=$~$0$; (3) we co-add the individual SFHs; (4) for each step in lookback time, we derive median (50\% of the distribution, SFH-P50) and confidence ranges (16\% and 84\% of the distribution, SFH-P16 and SFH-P84, respectively) of the co-added SFHs. We then characterize these composite SFHs using the milestones t$_{0}$, t$_{10}$, and t$_{50}$ as described in Section 3. We derive uncertainties on the milestones using the bootstrapping method.\footnote[3]{For each sample, we build 10$^{3}$ bootstrap samples (of the same size of the original) of SFHs randomly selecting objects and allowing for repetitions.} We report in Table~2 the values of the milestones and the associated standard deviations. The two left panels in Figure~3 show for each sample the SFH-P50 (black solid line), SFH-P16 (lower limit of the shaded area) and SFH-P84 (upper limit of the shaded area). The right panels in Figure~3 present the analogous plot for the composite sSFR histories (sSFRH) defined by the median, 16\,th, and 84\,th percentiles of the distribution of sSFR at each lookback time step (sSFRH-P50, sSFRH-P16, and sSFRH-P84 respectively). Each panel also shows the $t_{0}$, $t_{10}$, and $t_{50}$ derived from the corresponding median SFH-P50 ($t_{0,\mathrm{SFH-P50}}$, $t_{10,\mathrm{SFH-P50}}$, and $t_{50,\mathrm{SFH-P50}}$).
 
Our LMSFGs present a short median SFH that forms 90\% of the median mass in the bin (log$M_{*}/M_{\sun}\sim7.7$) in the 0.5--1.8~Gyr ($t_{10, \mathrm{SFH-P16}}$--$t_{10, \mathrm{SFH-P84}}$) period prior the observation, and 50\% in the last 0.3--0.5~Gyr ($t_{50, \mathrm{SFH-P16}}$--$t_{50, \mathrm{SFH-P84}}$). These results reinforce the idea of a recent stellar mass formation of LMSFGs, consistent with what \cite{2012ApJ...745..149L} obtained for SDSS star forming galaxies through the SED-fitting analysis and using an MSI approach. Recent formation of low-mass galaxies also matches the downsizing cosmological trend of galaxy formation \citep{1996AJ....112..839C}. \emph{Recent formation} refers to the individual time reference system of each galaxy. Early SF activity is not dismissed given the large dispersion in $t_{0,\mathrm{SFH-P50}}$. 

We find agreement (within the dispersion) between the timescales of the median SFHs of the galaxies in the two stellar mass bins. This is probably due to the small difference between the ranges of stellar mass (the difference between the medians is $0.7$~dex). Nonetheless, the three milestones tend to be larger for higher-mass galaxies.

The values of SFR along the whole median SFHs remain lower for the LMSFGs than for the secondary sample. Both composite SFHs show a statistically significant jump at $t=0$, indicative of the prominence of a current episode of enhanced SF. We note that the strong bursts are not caused by biases in the model library. The sSFR prior adopted to fit the observations is flat in the range $-2~<~$log\,sSFR/Gyr$^{-1}$~$<$~2 (smoothly declining on the edges). Similar individual SFHs have been obtained already by \cite{2014ApJ...789...96A} using our same approach \citep{2012MNRAS.421.2002P} for low-mass systems selected by their strong emission lines.

The median sSFRHs present clear differences between the two mass bins. Higher-mass galaxies tend to present higher sSFR than low-mass galaxies at early times and lower values than low-mass galaxies at the time they are observed. The shape of the median sSFRH of higher-mass galaxies looks roughly flat with an abrupt start, whereas it is bell shaped for the LMSFGs sample reaching a maximum at lookback times $\sim$\,1~$\mathrm{Gyr}$.

We note that the majority of the galaxies in the observed samples are characterized by a high sSFR. In such cases, the most recent burst of SF can outshine the older stellar populations, especially at ultraviolet and optical rest-frame wavelengths. We include photometry at larger wavelengths to better constrain the emission by more evolved stars (older than $\sim$2~Gyr). Furthermore, to evaluate our ability to reconstruct the SFHs given our data and library of SF and metal enrichment histories, we perform the following test. Using the same procedure described in Section~3, we fit the synthetic photometry of the best-fitting model of each of our targets assuming the same photometric uncertainties as in the  data and excluding such model from the library. In other words, we apply our methodology to synthetic galaxies for which we previously know the values of $t_{0}$, $t_{10}$, and $t_{50}$. The retrieved $t_{10}$ and $t_{50}$ present in general good agreement with the real values for any timescale with median deviations in Gyr (16\,th and 84\,th percentiles) of 0.2 ($-0.1$, 0.4), and 0.0 ($-0.2$, 0.2) respectively. Our methodology tends to overestimate $t_{0}$ by 1.2 (0.5, 2.3) Gyr. This happens because the derived SFHs can include masses that could have been formed at early stages without leaving any trace in the photometry.

\section{Summary and Conclusions}

We have analyzed a sample of 31~LMSFGs detected at 0.3~$<$~$z_{\mathrm{spec}}$~$<$~0.9 and with stellar masses 7.3~$\le$~$\mathrm{log}M_{*}/M_{\sun}$~$\le$~8.0, and a secondary sample of 43~spectroscopically confirmed more massive galaxies (8.0~$<$~$\mathrm{log}M_{*}/M_{\sun}$~$\le$~9.1) in the same redshift range. We have used the tool developed by \cite{2012MNRAS.421.2002P}, which combines physically motivated SF and chemical enrichment histories from cosmological simulations, with state-of-the-art models of stellar population synthesis, nebular emission, and attenuation by dust to constrain their SFHs. This approach allows us to perform the SED analysis including photometric (broadband/medium-band) and spectroscopic (EWs of emission lines) information. 
The main conclusions of our study are as follows:
   \begin{enumerate}
      \item The median SFH of our sample of LMSFGs suggests that $90\%$ of the stellar mass is formed in a 0.5--1.8~Gyr ($t_{10, \mathrm{SFH-P16}}$--$t_{10, \mathrm{SFH-P84}}$) period prior the observation. Our results reinforce the idea of a recent stellar-mass formation for LMSFGs at intermediate redshifts. They are consistent with the previous work about SFHs of star forming galaxies carried out by \cite{2012ApJ...745..149L} and with the downsizing cosmological frame \citep{1996AJ....112..839C}. 
      \item The estimated SFRs and stellar masses of the galaxies in both final samples are consistent with the star forming MS over 2~dex in stellar mass \citep[e.g.,][]{2007ApJ...660L..43N,2012ApJ...754L..29W}. 
      \item We find good agreement in the SFH timescales across the whole stellar mass range 7.3~$\le$~$\mathrm{log}M_{*}/M_{\sun}$~$\le$~9.1.
   \end{enumerate}

\acknowledgments
We are grateful to the anonymous referee for a very thorough report
which led to a substantial improvement of the paper.
We acknowledge support from the Spanish Programa Nacional de Astronom\'ia y Astrof\'isica: Project AYA2009-10368 and AYA2012-30717.
This work has used the Rainbow Cosmological Surveys Database, which is operated by the Universidad Complutense de Madrid (UCM), partnered with the University of California Observatories at Santa Cruz (UCO/Lick, UCSC). 
L.R.M. thanks Pablo G. P\'{e}rez-Gonz\'{a}lez for his technical advice on Rainbow usage and enriching discussions, and Roger Griffith for providing the latest version of the morphological catalogs on the E-CDF-S field.
C.P. acknowledges fundings by the KASI-Yonsei Joint Research Program for the Frontiers of Astronomy and Space Science funded by the Korea Astronomy and Space Science Institute. S.C. acknowledges support from the European Research Council via an advanced grant under grant agreement No. 321323-NEOGAL.
{\it Facility:} \facility{VLT:Melipal}.


\end{document}